\documentclass[onecolumn]{aastex631}

\usepackage{physics}
\usepackage{soul}
\usepackage{CJK}
\usepackage{hyperref}

\newcommand{\mstar}{\ensuremath{M_\star}}
\newcommand{\nmad}{\ensuremath{\sigma_{\rm NMAD}}}

\begin{document}

\begin{CJK*}{UTF8}{gbsn}
\title{Stellar Masses and Star-Formation Rates of Galaxies and AGNs in the eFEDS GAMA09 Field}

\author[0000-0002-6990-9058]{Zhibo Yu (喻知博)}
\affiliation{Department of Astronomy and Astrophysics, 525 Davey Lab, The Pennsylvania State University, University Park, PA 16802, USA}
\email{Email: zvy5225@psu.edu}
\affiliation{Institute for Gravitation and the Cosmos, The Pennsylvania State University, University Park, PA 16802, USA}
\author[0000-0002-4436-6923]{Fan Zou (邹凡)}
\affiliation{Department of Astronomy and Astrophysics, 525 Davey Lab, The Pennsylvania State University, University Park, PA 16802, USA}
\affiliation{Institute for Gravitation and the Cosmos, The Pennsylvania State University, University Park, PA 16802, USA}
\author[0000-0002-0167-2453]{William N. Brandt}
\affiliation{Department of Astronomy and Astrophysics, 525 Davey Lab, The Pennsylvania State University, University Park, PA 16802, USA}
\affiliation{Institute for Gravitation and the Cosmos, The Pennsylvania State University, University Park, PA 16802, USA}
\affiliation{Department of Physics, 104 Davey Laboratory, The Pennsylvania State University, University Park, PA 16802, USA}



\begin{abstract}

The eFEDS is a wide $\approx140\,\deg^2$ field that has extensive multiwavelength coverage. To improve the utility of the existing data, we use \texttt{CIGALE} to fit source Spectral Energy Distributions (SEDs) from X-rays to far-infrared (FIR) mainly to derive stellar masses (\mstar) and star-formation rates (SFRs) for normal galaxies and X-ray Active Galactic Nuclei (AGNs). The catalog consists of 2,057,027 galaxies and 10,373 X-ray AGNs located in the $\approx60\deg^2$ GAMA09 sub-field. Comparing our \mstar\ with other available catalogs and our SFRs with FIR-derived SFRs, we demonstrate the general reliability of our SED-fitting measurements. Our catalog is publicly available at \href{https://doi.org/10.5281/zenodo.10127224}{10.5281/zenodo.10127224}.

\end{abstract}


\section{Introduction} \label{sec:intro}

The eROSITA Final Equatorial Depth Survey (eFEDS) was the largest observational investment during the eROSITA performance verification phase. The entire field, spanning $\approx140\,\deg^2$, was observed to a depth of $\approx2.2\,\mathrm{ks}$ by eROSITA \citep{Brunner+2022,Salvato+2022}. 
eFEDS was constructed to encompass the $\approx60\deg^2$ GAMA09 field \citep{Driver+2022}, which has rich multi-wavelength data. Given the X-ray coverage from eROSITA and the well-cataloged photometric data from UV to FIR on a $10^2$-$\deg^2$ scale, eFEDS is useful for both AGN and galaxy studies. To facilitate future studies, we report \mstar\ and SFRs for $\approx\,2$ million sources in the eFEDS GAMA09 field primarily utilizing the GAMA-derived photometric catalog that is based upon KiDS and VIKING imaging data \citep{Bellstedt+2020}. The SED fitting was performed with \texttt{CIGALE} \citep{Boquien+2019,Yang+2022}.

\section{Data and Methods}\label{sec::method}

Our analysis focuses on the GAMA09 region since the remaining parts of eFEDS lack sufficient multi-wavelength coverage, especially the NIR coverage provided by KiDS and VIKING, which are crucial in deriving reliable \mstar. We select X-ray AGNs based on the eFEDS X-ray main catalog \citep{Brunner+2022}, and the intrinsic X-ray fluxes are taken from \citet{Liu+2022} who corrected for absorption via X-ray spectral analyses. The rest of the sources are classified as galaxies.  The UV-to-FIR photometry is from the GAMA-KiDS-VIKING (GKV) catalog compiled in \citet{Bellstedt+2020}. In general, the multiwavelength coverage is uniform, reaching $5 \sigma$ depths of 24.2 mag and 21.3 mag for $i$-band and $K_S$-band, respectively. We also incorporate the HSC-Wide survey to support the optical coverage, which uniformly covers the entire GAMA09 region with a $5\sigma$ limiting magnitude of 26.1 mag in $i$-band \citep{Aihara+2022}. The effective filter response and zero-point calibration across different HSC bands are also corrected. Both the GKV and HSC-Wide catalogs have accounted for Galactic extinction. Additionally, we include \textit{Herschel} FIR photometry from the HELP collaboration \citep{Shirley+2021}. The photometric redshifts (photo-zs) and spectroscopic redshifts (spec-zs) are taken from the compilations of \citet{Salvato+2022} and \citet{Driver+2022} for X-ray AGNs and normal galaxies, respectively. The photo-zs generally have good quality with typical dispersions of a few percent and outlier fractions of less than 10\%. We also drop sources near bright stars so that the impact on our overall photometry is minimal. Our final sample contains 2,057,027 normal galaxies and 10,373 X-ray AGNs \citep[including 9,856 with primary counterparts and 517 with secondary counterparts based upon the identifications of][]{Salvato+2022}.

We apply the same methods as in \citet{Zou+2022} to derive the best-fit \mstar\ and SFRs. Briefly, \texttt{CIGALE} assumes an energy balance principle and decomposes a SED into several user-defined components (including AGNs). For normal galaxies and X-ray AGNs, we use the dense-grid \texttt{CIGALE} parameter settings for normal galaxies and AGN candidates in \citet{Zou+2022}, respectively (see their Tables 4 and 5).
We fit the near-UV (NUV) to FIR SEDs for galaxies and add X-ray photometry for AGNs. To account for systematic uncertainties, 0.05 mag error is added in quadrature for all bands from NUV to NIR.

\section{Results}\label{sec::results}

\begin{figure}[t!]
    \centering
    \includegraphics[width=\columnwidth]{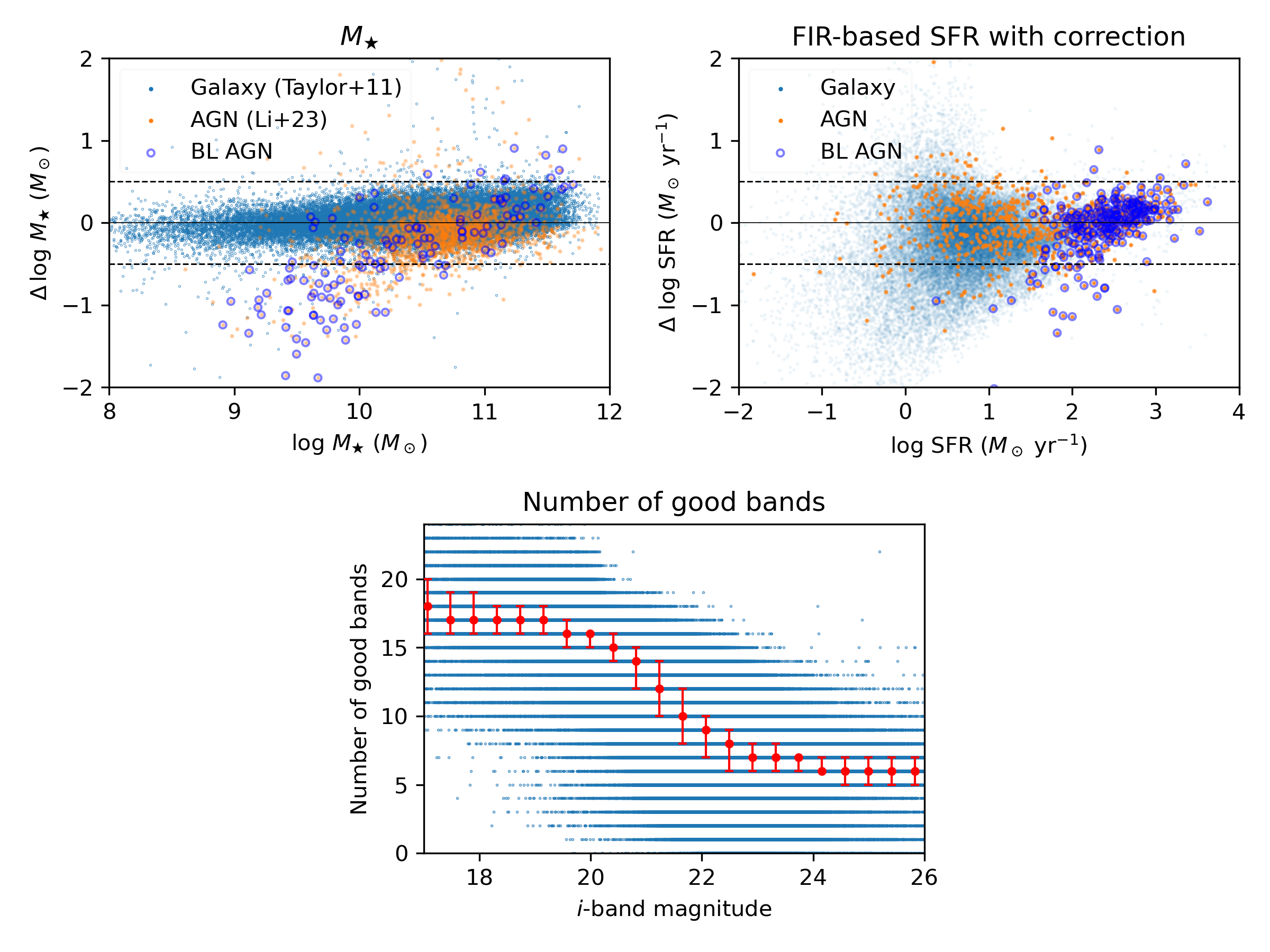}
    \caption{Top-left: Comparison between our \mstar\ and \mstar\ from existing catalogs in eFEDS. Top-right: Comparison between our SED-based SFRs and FIR-based SFRs. The dashed lines indicate 0.5 dex deviations. Note that we only plot sources with $i$-mag $<22$. Bottom: Number of good bands (with $\rm SNR > 5$) vs. $i$-mag. Each blue point represents a source, and the red error bars represent the median, 25th, and 75th percentile of the number of good bands in each magnitude bin.}\label{fig:sed}
\end{figure}

To evaluate the reliability of our \mstar\ measurements, we compare our \mstar\ for normal galaxies with the results from the GAMA collaboration ($M_{\star,\rm ref}$) \citep{Taylor+2011}, which only include bright sources with GAMA spectra. For X-ray AGNs, we refer to the \mstar\ measurements from \citet{Li+2023}. The comparisons are shown in the top-left panel of Figure~\ref{fig:sed} where $\Delta\log\mstar=\log\mstar-\log M_{\star,\rm ref}$. For 59,653 normal galaxies and 2,029 X-ray AGNs, $\nmad=0.12$ and $0.23$ with median $\Delta\log\mstar=0.02$ and $-0.12$, where \nmad\ is the normalized median absolute deviation (NMAD).\footnote{NMAD is defined as 1.4826 $\times$ median absolute deviation.} To assess the SFR measurements, we compare SFRs based upon our SED-fitting and FIR-based SFRs ($\rm SFR_{FIR}$) that are derived following the method in \citet{Chen+2013}. We also correct for old-star heating following Equation~25 in \citet{Zou+2022}. Figure~\ref{fig:sed} top-right panel shows the comparisons between different SFRs for both galaxies and X-ray AGNs, where $\Delta\log\mathrm{SFR}=\log\mathrm{SFR}-\log\mathrm{SFR_{FIR}}$. For 34,610  galaxies and 862 X-ray AGNs with FIR signal-to-noise ratio (SNR) $> 5$, $\nmad=0.41$ and $0.28$ with median $\Delta\log\mathrm{SFR}=-0.19$ and $0.04$, respectively. These values are generally as good as those in \citet{Zou+2022} where they fit SEDs to three million sources in the $\approx13.2\deg^2$ XMM-SERVS fields. Due to the large area and shallow X-ray depth of eFEDS, there will be a significant fraction of BL AGNs whose AGN components generally dominate the NIR emission, causing less reliable $\mstar\ $ measurements. Thus, we also plot Broad-Line AGN (BL AGN) candidates in \mbox{Figure~\ref{fig:sed}}, which are selected as having AGN components constituting $>50\%$ of the total flux density at rest-frame 1 $\mu$m. The impact of a high AGN contribution on $\mstar\ $ is clearly shown by the widely scattered $\Delta\log\mstar$ ($\nmad=0.66$). Apart from eFEDS, we also apply the same method for SED-fitting to COSMOS, and the results are consistent with the COSMOS2020 catalog by \citet{Weaver+2022}.

We further estimate the nominal depth of our measurements. We define ``good bands" as those with $\rm SNR >5$, and we plot the number of good bands vs. $i$-band magnitude in Figure~\ref{fig:sed} bottom panel. The plot indicates that the SED quality in eFEDS degrades at $i$-mag~$\approx 21.5$. Approximately 30\% of our sources are brighter than this magnitude. Among these, 10\% of normal galaxies and 43\% of X-ray AGNs have available spec-zs. We only show sources with $i$-mag $< 22$ in Figure~\ref{fig:sed} top panels, which constitute $\gtrsim90\%$ of the sources that we compared. We warn readers to use our catalog at $i$-mag $\gtrsim 22$ cautiously, as fainter sources normally contain only $\sim6$ bands primarily from HSC, and it is harder to constrain their properties due to lack of NIR coverage.

Our catalog is available at \href{https://doi.org/10.5281/zenodo.10127224}{10.5281/zenodo.10127224}. We provide the source \mstar, SFRs, classifications (X-ray AGNs vs. normal galaxies), and necessary information that can be helpful to cross-reference GAMA objects and/or eFEDS X-ray/optical-IR counterparts. We also provide AGN fractional contributions at rest-frame $5000$ \r{A}, 1 $\mu$m, and integrated $8-1000$ $\mu$m for X-ray AGNs to help readers identify sources with less-reliable $\mstar$.
Readers should also note that our AGN selection is based upon X-ray detection only, which is generally a good tracer of black-hole accretion rates (BHAR), but this selection is incomplete. To reach better completeness, one should combine multiple selection methods such as mid-infrared \citep[MIR; e.g.,][]{Zou+2022} and radio \citep[e.g.,][]{Zhu+2023}.


{\bf Acknowledgments}

We acknowledge support from NSF grant AST-2106990 and Penn State.

%

\vspace{5mm}


\newpage

\appendix
\section{SEDs in COSMOS}

The COSMOS field is one of the LSST Deep Drilling Fields (DDFs) and has been well characterized by many past studies \citep[e.g.,][]{Marchesi+2016,Weaver+2022}. Our purpose in working on COSMOS is to consistently measure the galaxy properties via SED-fitting in the same manner as for other DDFs \citep[e.g.,][]{Zou+2022}. In this work, our methods for COSMOS are the same as those for eFEDS and \citet{Zou+2022}. We focus on the 1.27 deg$^2$ UltraVISTA footprint inside COSMOS because sufficiently deep NIR coverage is critical in deriving reliable \mstar. The final catalog consists of 709,087 normal galaxies and 2,209 X-ray AGNs. The absorption-corrected X-ray data are from \citet{Civano+2016} and \citet{Marchesi+2016}. The UV-to-MIR data are from \textit{the Farmer} catalog in the COSMOS2020 data release \citep{Weaver+2022}. To optimize the magnitude offsets in each band, we apply the correction in Table~3 of \citet{Weaver+2022}. We also add the 24--500 $\mu$m data from the ``super-deblended" FIR photometric catalog in \citet{Jin+2018}. The Galactic extinction is corrected following the methods in Section 2.6 of \citet{Zou+2022}.  The redshifts for X-ray AGNs and normal galaxies are from \citet{Marchesi+2016} and \citet{Weaver+2022}, respectively. The quality of the photo-zs is generally good, with typical dispersions of 1--4\% and outlier fractions of a few percent.

\begin{figure}[ht!]
    \centering
    \includegraphics[width=\columnwidth]{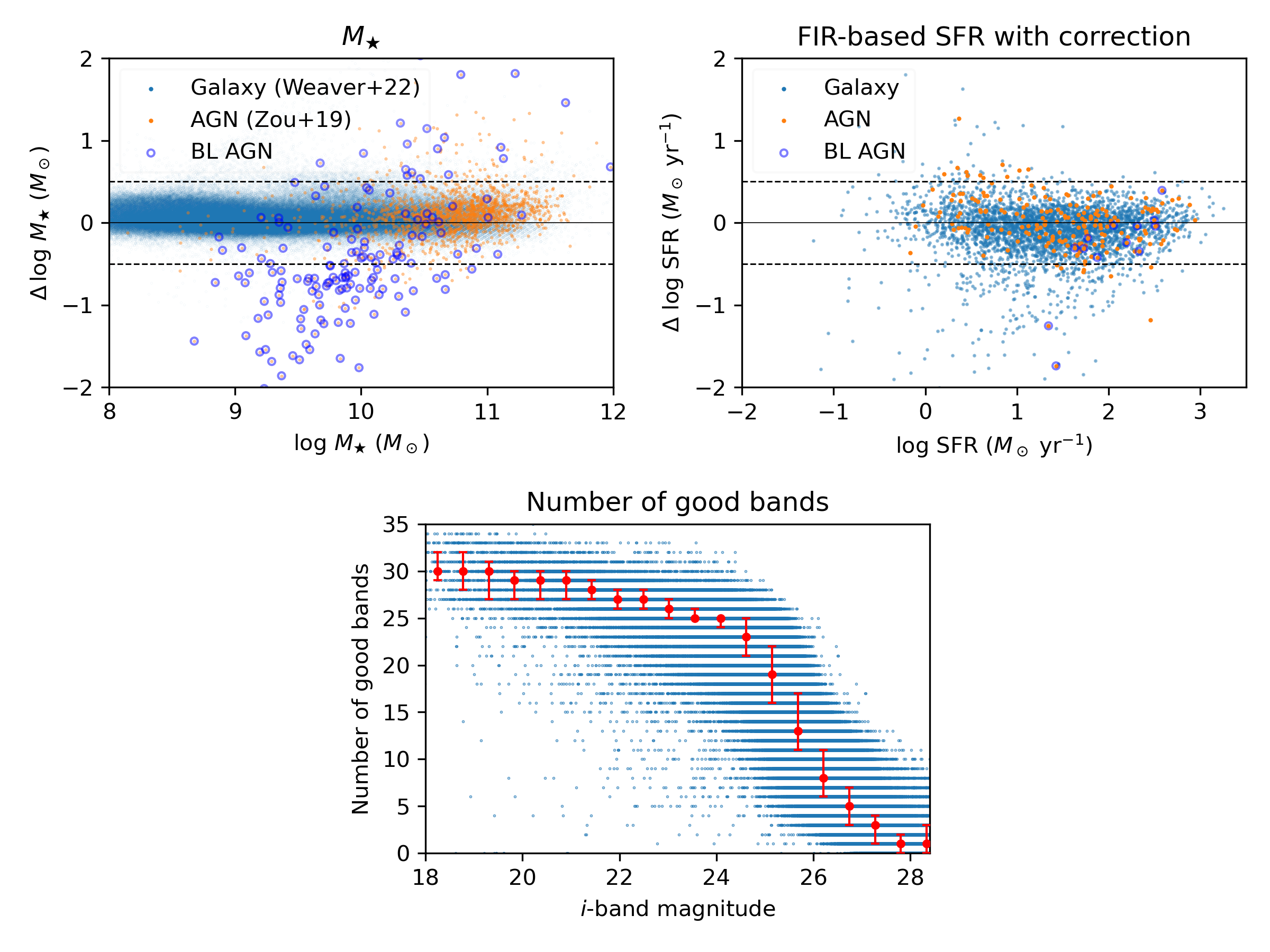}
    \caption{Top-left: Comparison between our \mstar\ and \mstar\ from existing catalogs in COSMOS. Top-right: Comparison between our SED-based SFRs and FIR-based SFRs. The dashed lines indicate 0.5 dex deviations. Note that we only plot sources with $i$-mag $<26$. Bottom: Number of good bands (with $\rm SNR > 5$) vs. $i$-mag. Each blue point represents a source, and the red error bars represent the median, 25th, and 75th percentile of the number of good bands in each magnitude bin.}\label{fig:sed-cosmos}
\end{figure}

We compare our \mstar\ with other measurements in the top-left panel of Figure~\ref{fig:sed-cosmos}. The \mstar\ values for 703,081 galaxies are compared with those in \textit{the Farmer} catalog. The median $\Delta\log \mstar=0.07$, and $\nmad=0.15$. We compare 2,086 X-ray AGNs with those in \citet{Zou+2019}, who also include AGN components when deriving \mstar. The median $\Delta\log \mstar=0.09$, and $\nmad=0.21$. We also select BL AGN candidates with AGN components contributing $>50\%$ of the total flux density at rest-frame 1 $\mu$m. The result indicates this criterion is efficient in selecting AGNs with problematic $\mstar$ measurements as almost all sources with $\Delta\log\mstar<-0.5$ are selected. The impact of these BL AGN candidates on COSMOS should be much reduced compared to eFEDS because of the small area and greater depth of COSMOS. This is supported by the much-reduced fraction of sources with AGN contribution at rest-frame 1 $\mu$m $>50\%$ in COSMOS (9\%) than in eFEDS (32\%). In addition, among our X-ray AGNs, 36 are spectroscopically confirmed BL AGNs by \citet{Suh+2020}. The median $\Delta\log\mstar$ is $-0.40$, which is consistent with the findings in \citet{Zhuang+2023}. In Figure~\ref{fig:sed-cosmos} top-right panel, we compare SED-based SFRs and FIR-based SFRs corrected for old-star heating. For 4,017 galaxies and 225 AGNs with FIR SNR $>5$, $\nmad=0.23$ and $0.27$ with median $\Delta\log\rm SFR=-0.06$ and $0.05$, respectively. Our measurements in COSMOS are generally consistent with the previous results. 

The nominal depth in COSMOS is also shown in the bottom panel of Figure~\ref{fig:sed-cosmos}. The number of good bands in COSMOS is much larger than that in eFEDS at the bright end, but it dramatically degrades at $i$-mag $\approx25.5$. Approximately 35\% of our sources are brighter than this magnitude. Among these, 64\% of X-ray AGNs have spec-zs from \citet{Marchesi+2016}.\footnote{The spec-zs for normal galaxies in the COSMOS2020 data release are not publicly available, so we only adopt their photo-zs. Since their photo-zs agree well with the spec-zs, the adoption of photo-zs should not materially affect our conclusions.} In the Figure~\ref{fig:sed-cosmos} top panels, we only show sources with $i$-mag $<26$. We warn readers to use our catalog at $i$-mag $\gtrsim26$ cautiously as fainter sources generally have only a handful of bands, so the SED fitting becomes unreliable.

\newpage

\bibliography{references}{}
\bibliographystyle{aasjournal}


\end{CJK*}
\end{document}